\documentclass[preprintnumbers,amsmath,amssymb,floatfix,showpacs]{revtex4}

\usepackage{epsf}
\usepackage[hang,nooneline]{subfigure}
\usepackage{natbib}
\usepackage{graphicx}

\usepackage{aas_macros}

\begin{document}

\title{Axial quasi-normal modes of Einstein-Gauss-Bonnet-dilaton neutron stars}

\author{
{\bf Jose Luis Bl\'azquez-Salcedo$^1$},
{\bf Luis Manuel Gonz\'alez-Romero$^2$}\\
{\bf Jutta Kunz$^1$},
{\bf Sindy Mojica$^1$},
{\bf Francisco Navarro-L\'erida$^3$}\\
\vspace{0.5truecm}
$^1$
 Institut f\"ur  Physik, Universit\"at Oldenburg\\ Postfach 2503,
D-26111 Oldenburg, Germany\\
$^2$
Departamento de F\'{\i}sica Te\'orica II, Ciencias F\'{\i}sicas\\
Universidad Complutense de Madrid, E-28040 Madrid, Spain\\
$^3$
Departamento de F\'{\i}sica At\'omica, Molecular y Nuclear, Ciencias F\'{\i}sicas\\
Universidad Complutense de Madrid, E-28040 Madrid, Spain
}
\vspace{0.5truecm}

\vspace{0.5truecm}

\vspace{0.5truecm}
\date{
\today}

\begin{abstract}
We investigate axial quasi-normal modes of realistic neutron stars in Einstein-Gauss-Bonnet-dilaton gravity. We consider 8 realistic equations of state containing nuclear, hyperonic, and hybrid matter. We focus on the fundamental curvature mode and compare the results with those of pure Einstein theory. We observe that the frequency of the modes is increased by the presence of the Gauss-Bonnet-dilaton, while the impact on the damping time is typically smaller. Interestingly, we obtain that universal relations valid in pure Einstein theory still hold for Einstein-Gauss-Bonnet-dilaton gravity, and we propose a method to use these phenomenological relations to constrain the value of the Gauss-Bonnet coupling.
\end{abstract}

\maketitle

\section{Introduction}

Neutron stars can be used as laboratories for testing gravity and matter in the strong energy regime. In particular, neutron stars provide information about the structure of matter at high energies, they are sources of gravitational waves, and they also allow us to test modifications of Einstein gravity \cite{clifford:LRR2014}.

Regarding the properties of matter, the equation of state (EOS) of nuclear matter beyond nuclear density ($10^{15} g/cm^3$) is not known, since Earth-based laboratories cannot achieve these energies yet. But these densities are present in the inner regions of neutron stars, and hence observations of neutron stars and their properties can shed light into the behavior of matter at high energy. Most models of neutron stars predict a layer structure, essentially a crust (where the EOS is better understood and matter has a solid crystalline structure similar to a metal) enveloping the core \cite{haensel2006neutron}. Typically the models for the EOS at the core can include plain nuclear matter, hyperonic matter, quark condensates, or a mixture of these possible states. Different EOS predict different properties of the neutron stars. For example, depending on the stiffness of the EOS, different mass-radius relations are obtained. Nowadays, the two most massive pulsars measured are PSR J1614-2230, with a mass of $1.97\pm0.04 M_{\odot}$ \cite{2010Natur.467.1081D}, and PSR J0348+0432, with a mass of $2.01\pm0.04 M_{\odot}$ \cite{Antoniadis26042013}. These observations constrain the possible EOS. The EOS also modify the spectrum of resonant frequencies from gravitational wave emission. Hence the importance of considering realistic EOS for the composition of the neutron star. 

On the other hand, neutron stars are among major candidates as sources of gravitational waves. The typical frequencies of these objects lie inside the range of detectability of current large-scale interferometric gravitational wave detectors like LIGO, VIRGO and KAGRA \cite{Pitkin:LRR2011}. The resonant frequencies and damping times of gravitational waves produced by a compact object, can be studied using the quasi-normal mode (QNM) formalism \cite{lrr-1999-2,0264-9381-16-12-201,2008GReGr..40..945F}. The eigen-frequencies are given by a complex number, where the real part gives us the frequency of the resonance, and the imaginary part gives us the inverse of the damping time. The quasi-normal mode spectrum is composed of different families: \textit{axial} modes only couple to space-time oscillations, while \textit{polar} modes can couple to matter perturbations. 
Quasi-normal modes carry information about the composition of the neutron star. But even more, the spectrum could also contain information about possible modifications of Einstein theory of gravity.

A full theory of gravity should be compatible with quantum theory. Hence, General Relativity is expected to be modified at some scale, in particular, around the space-time of compact objects such as neutron stars and black holes. One of the most promising unified theories is String theory. Because of the complexity of studying the full string theory, analysis is often made in the low-energy regime. A simple particular case is Einstein-Gauss-Bonnet-dilaton (EGBd) theory in 4 dimensions, obtained from the low-energy regime of heterotic string theory. The action of this effective theory contains, in addition to the scalar of curvature, a Gauss-Bonnet modification of the action coupled to a real scalar field: the dilaton \cite{0264-9381-24-2-006}. The EGBd theory has nice features: for example, it does not present ghosts, the equations of motion are still second order, and wormholes without exotic fields can be produced \cite{2011PhRvL.107A1101K,2012PhRvD..85d4007K}

Astrophysically relevant objects have been recently studied in EGBd theory, as well as similar modified gravity theories \cite{2015arXiv150107274B}. Static dilatonic black holes in EGBd theory were first studied in \cite{1996PhRvD..54.5049K}. Slowly rotating black holes were considered in \cite{2009PhRvD..79h4031P,2011PhRvD..84h7501P}, axial QNM in EGBd black holes were obtained in \cite{2009PhRvD..79h4031P}, and generalizations of the Kerr black holes in EGBd theory were studied in \cite{2011PhRvL.106o1104K,2014PhRvD..90f1501K}.

Neutron stars in modified theories of gravity have been studied extensively. In EGBd gravity, neutron stars in the slow rotation approximation were studied in \cite{2011PhRvD..84j4035P}, while the multipole structure of fast rotating neutron stars were recently investigated in \cite{2014PhRvD..90f1501K}. 

In the context of neutrons stars, several universal relations between different parameters of the stars were recently found. For rotating neutrons stars in Einstein theory, I-Love-Q relations between the moment of inertia, Love numbers and quadrupole moment were extensively studied \cite{2013Sci...341..365Y}. In \cite{2014PhRvD..90f1501K} it was indicated that these I-Q relations also hold in EGBd theory.

Phenomenological relations between mass, radius and QNM spectrum were also studied in Einstein theory \cite{2001MNRAS.320..307K,2004PhRvD..70l4015B}, and more recently in \cite{2013PhRvD..87j4042B,2014PhRvD..89d4006B}. It is interesting to investigate whether the universal relations still hold in EGBd theory, and to find the possible implications once gravitational waves are detected.

In this paper we consider for the first time quasi-normal modes of realistic neutron stars in EGBd theory. We perform the analysis for 8 realistic state-of-the-art EOS, continuing with the analysis performed for Einstein theory in \cite{2013PhRvD..87j4042B,2014PhRvD..89d4006B}. We consider two EOS for nuclear matter (SLy and APR4), three for hyperonic matter (WCS1-2 and BHZBM), and another three for hybrid matter (ALF4, WSPHS3 and BS4). We calculate the spectrum of axial curvature modes for a large value of the GB coupling \cite{2012PhRvD..86h1504Y}. We compare the impact of the GBd term with the pure Einstein case, and look for observational implications.

In section \ref{sec_theory} we briefly present the general formalism for neutron stars in EGBd theory. In section \ref{sec_QNM} we present the QNM formalism for neutron stars in EGBd theory, we present the equations for axial perturbations, and we briefly introduce the numerical method used to calculate the modes for realistic EOS. In section \ref{sec_results} we discuss the numerical results comparing EGBd with Einstein gravity, and we present new universal relations for EGBd. Finally in section \ref{sec_conclusions} we present a summary and an outlook.

Let us start by introducing briefly the general theory of neutron stars in EGBd gravity.

\section{Neutron stars in Einstein-Gauss-Bonnet theory} \label{sec_theory}

\subsection{General theory}

Einstein-Gauss-Bonnet-dilaton theory in four dimensions is described by the following action \cite{2011PhRvL.106o1104K,2014PhRvD..90f1501K}:

\begin{equation}
I= \frac{1}{16\pi G} \int d^4x 
\sqrt{-g}\biggl[R - \frac{1}{2}(\partial_{\mu}\Phi)^2 + \alpha e^{-\beta\Phi}R_{GB}^2 + L_{matter}\biggr]
,
\end{equation}
where $\alpha$ is the Gauss-Bonnet coupling constant, $\beta$ is the dilaton coupling constant, $R$ is the curvature scalar, $\Phi$ is the dilaton field, and $R_{GB}^2$ is the Gauss-Bonnet term
\begin{equation}
R_{GB}^2 = R_{\alpha\beta\gamma\delta}R^{\alpha\beta\gamma\delta} -4R_{\alpha\beta}R^{\alpha\beta} + R^2.
\end{equation}

The field equations are obtained from variations of this action and yield the modified Einstein equations
\begin{equation}
G_{\mu\nu} = \frac{1}{2}\biggl[\nabla_{\mu}\Phi \nabla_{\nu}\Phi -\frac{1}{2}g_{\mu\nu}\nabla_{\alpha}\Phi \nabla^{\alpha}\Phi \biggr] - \alpha e^{-\beta\Phi}\biggl[H_{\mu\nu} + 4(\beta^2\nabla^{\alpha}\Phi\nabla^{\beta}\Phi -\beta\nabla^{\alpha}\Phi\nabla^{\beta}\Phi)P_{\mu\alpha\nu\beta}  \biggr] + 8\pi T_{\mu\nu},
\end{equation}
where $G_{\mu\nu}$ is the Einstein tensor and
\begin{eqnarray}
H_{\mu\nu}=2(RR_{\mu\nu}-2R_{\mu\alpha}R^{\alpha}_{\ \nu}-2R_{\mu\alpha\beta\nu}R^{\alpha\beta}+R_{\mu\alpha\beta\delta}R_{\nu}^{\ \alpha\beta\delta})-\frac{1}{2}g_{\mu\nu}R^2_{GB}, \\
P_{\alpha\beta\gamma\delta}=R_{\alpha\beta\delta\gamma}+g_{\alpha\gamma}R_{\delta\beta}-g_{\alpha\delta}R_{\gamma\beta}+g_{\beta\delta}R_{\gamma\alpha}-g_{\beta\gamma}R_{\delta\alpha}+\frac{1}{2}Rg_{\alpha\delta}g_{\gamma\beta}-\frac{1}{2}Rg_{\alpha\gamma}g_{\delta\beta}.
\end{eqnarray}
In four dimensions the $H_{\mu\nu}$ tensor vanishes.

The matter inside 
the star is considered to be a perfect fluid with stress-energy tensor
\begin{equation}
T_{\mu \nu } = \left( p + \rho \right)u_{\mu}u_{\nu} - p g_{\mu \nu},
\end{equation} 
where $p$ is the pressure, $\rho$ is the energy density and $u$ is the 4-velocity.

In addition we obtain the dilaton field equation
\begin{equation}
\nabla^2 \Phi=\alpha \beta e^{-\beta\Phi}R_{GB}^2.
\end{equation} 
Note that the value of the dilaton coupling constant obtained from string theory is $\beta=1$.

\subsection{Static and spherically symmetric neutron stars}

We consider static neutron stars described by a static and spherically symmetric space-time with metric ansatz
\begin{equation}
ds^{2}=g^{(0)}_{\mu\nu}dx^{\mu}dx^{\nu}=e^{2\nu}dt^{2}-e^{2\lambda}dr^{2} - r^{2}(d\theta^{2} 
+ \sin^{2}\theta d\varphi^{2}),
\end{equation} 
where the functions $\nu$ and $\lambda$ are just functions of the radial coordinate $r$. The dilaton field is described by the function $\Phi=\Phi(r)$. The mass function $m$ is defined as $e^{-2\lambda}=1-\frac{2m}{r}$.

The resulting generalized Tolman-Oppenheimer-Volkoff (TOV) equations in EGBd theory can be written as \cite{2011PhRvD..84j4035P}
:
\begin{eqnarray} 
\frac{d\nu}{dr}&=&
\frac{1}{8}\frac{r e^{\beta\Phi}\left[r^2\left({\frac{d\Phi}{dr}}\right)^{2}(r-2m)+8m+32\pi r^3 p\right]}{(r-2m)\left[4\alpha\beta\frac{d\Phi}{dr}(3m-r)+e^{\beta\Phi}r^2\right]}
, \label{TOV_EGBd_nu} \\
\frac{dp}{dr}&=&
-\frac{d\nu}{dr}(\rho+p)
, \label{TOV_EGBd_p}  \\
\frac{dm}{dr}&=&
\biggl[
32\pi\rho r^5 e^{\beta\Phi} -  (r-2m)\left({\frac{d\Phi}{dr}}\right)^{2}\left(32\alpha\beta^2 mr - e^{\beta\Phi}r^4 \right)
\nonumber \\ &+& 32\alpha\beta rm \frac{d^2\Phi}{dr^2}(r-2m) + 32\alpha\beta m \frac{d\Phi}{dr} (3m-r)
\biggr] \nonumber \\ &\times&
\left[
8r\left(
4\alpha\beta \frac{d\Phi}{dr}(3m-r) + e^{\beta\Phi}r^2 
\right)
\right]^{-1}
. 
\label{TOV_EGBd_m}
\end{eqnarray} 
They reduce to the usual TOV equations when $\alpha=\beta=0$.

For the dilaton field we obtain the following equation:
\begin{eqnarray}
&& \frac{d^2\Phi}{dr^2}e^{\beta\Phi}r^4(r-2m) -
\frac{d\Phi}{dr}e^{\beta\Phi} r^3\left[3m-2r+r\frac{dm}{dr}-r(r-2m)\frac{d\nu}{dr}\right] \nonumber \\ &-&
\frac{d^2\nu}{dr^2} 16\alpha\beta mr (r-2m) - \frac{d\nu}{dr}16\alpha\beta \left[
rm(r-2m)\frac{d\nu}{dr}-r(3m-r)\frac{dm}{dr} +m(3m-r)
\right]
=0.  \label{dil_EGBd}
\end{eqnarray}

The space-time of a neutron star is described by two different regions. Inside the neutron star we have a non-vanishing pressure $p$ and density $\rho$, related by some equation of state. These functions become null at some $r=R$, where $R$ is the radius of the neutron star. For $r>R$ we have vanishing pressure and energy density.  

Regularity at the origin requires
\begin{equation}
p = p_c + O(r^2), \ \nu = \nu_c + O(r^2), \ m = O(r^3),  \ \Phi = \Phi_c + O(r^2). \label{bc_org}
\end{equation}

Asymptotic flatness requires

\begin{eqnarray}
\nu = O(r^{-1}), \ 
m = M + O(r^{-1}), \ 
\Phi = O(r^{-1}), \label{bc_inf}
\end{eqnarray}
where M is the total mass of the star, and in EGBd, does not coincide with $m(R)$, since the dilaton does not vanish for $r>R$.

\section{Non-radial perturbations for Einstein-Gauss-Bonnet-dilaton neutron stars} \label{sec_QNM}

\subsection{General formalism}

Following \cite{PhysRev.108.1063,PhysRevLett.24.737}, we consider linear non-radial perturbations to the metric and fluid, but now allowing also for possible perturbations of the dilaton field:

\begin{eqnarray}
g_{\mu\nu} = g_{\mu\nu}^{(0)} + \epsilon h_{\mu\nu} \\
p = p^{(0)} + \epsilon \delta p\\
\rho = \rho^{(0)} + \epsilon \delta \rho\\
u_{\mu} = u_{\mu}^{(0)} + \epsilon \delta u_{\mu}\\
\Phi = \Phi^{(0)} + \epsilon \delta \Phi,
\end{eqnarray}
where $\epsilon<<1$ is the perturbation parameter. At zeroth order we have the static and spherical solution given by the metric $g_{\mu\nu}^{(0)}$, the pressure $p^{(0)}$, the density $\rho^{(0)}$, the fluid 4-velocity $u_{\mu}^{(0)}$, and  the dilaton field $\Phi^{(0)}$. These functions satisfy equations (\ref{TOV_EGBd_nu})-(\ref{dil_EGBd}). At first order in the perturbation parameter $\epsilon$ in general we have the perturbation to the metric $h_{\mu\nu}$, pressure $\delta p$, density $\delta \rho$, 4-velocity $\delta u_{\mu}$ and dilaton $\delta \Phi$.

In general the perturbations depend non-trivially on all the coordinates. For the temporal dependence, it is convenient to simplify the temporal dependence by performing a Laplace transformation. For the angular part, we can expand in tensorial spherical harmonics \cite{RevModPhys.52.299}, introducing the rotational numbers $l,m$. Then the perturbations split into two decoupled channels: axial and polar perturbations. 

The axial perturbations transform as $(-1)^{l+1}$ under reflections of the angular coordinates, and do not couple to scalar perturbations. They are described by the metric perturbation
\begin{equation}
h_{\mu\nu}^{(axial)} =  \sum\limits_{l,m}\int d\omega\left[
\begin{array}{c c c c}
	0 & 0 & h_0	\frac{1}{\sin\theta}\frac{\partial}{\partial\phi}Y_{lm} & h_0	\sin\theta\frac{\partial}{\partial\theta}Y_{lm} \\
  0 & 0 & 0 & h_1	\sin\theta\frac{\partial}{\partial\theta}Y_{lm} \\
	h_0	\frac{1}{\sin\theta}\frac{\partial}{\partial\phi}Y_{lm} & 0 & 0 & 0 \\
	h_0	\sin\theta\frac{\partial}{\partial\theta}Y_{lm} & h_1	\sin\theta\frac{\partial}{\partial\theta}Y_{lm} & 0 & 0
\end{array}
\right]e^{-i\omega t}.
\end{equation}
We also have to consider the axial component of the 4-velocity perturbation
\begin{equation}
\delta u^{(axial)\mu} =  \sum\limits_{l,m}\int d\omega\frac{w_2}{\sin^2\theta}\left[0,0,{\sin\theta\frac{\partial}{\partial\phi}}Y_{lm},-\frac{\partial}{\partial\theta}Y_{lm}\right]e^{-i\omega t}
\end{equation}
The perturbation functions $(h_0, h_1, w_2)$ depend in general on the radial coordinate $r$, numbers $l$, $m$ and $\omega$.
 
The axial channel does not introduce perturbations to the pressure, energy density or dilaton. This only happens in the polar channel, which transforms as $(-1)^{l}$. The perturbations of the metric are then described by
\begin{equation}
h_{\mu\nu}^{(polar)} =  \sum\limits_{l,m}\int d\omega\left[
\begin{array}{c c c c}
	2Ne^{2\nu} & -H_1 & 0 & 0 \\
  -H_1 & -2Le^{2\lambda} & 0 & 0 \\
	0 & 0 & -2Te^{2\lambda} & 0 \\
	0 & 0 & 0 & -2Te^{2\lambda}
\end{array}
\right]Y_{lm}e^{-i\omega t}.
\end{equation}
The polar component of the 4-velocity perturbation is
\begin{equation}
\delta u^{(polar)\mu} =  \sum\limits_{l,m}\int d\omega\left[0,W Y_{l,m},V\frac{\partial}{\partial\theta}Y_{lm},\frac{1}{\sin^2\theta}V\frac{\partial}{\partial\phi}Y_{lm}\right]e^{-i\omega t}.
\end{equation}
The pressure, energy density and dilaton perturbations are just
\begin{eqnarray}
\delta p^{(polar)} =  \sum\limits_{l,m}\int d\omega p_1 Y_{l,m} e^{-i\omega t}, \\
\delta \rho^{(polar)} =  \sum\limits_{l,m}\int d\omega \rho_1 Y_{l,m} e^{-i\omega t}, \\
\delta \Phi^{(polar)} =  \sum\limits_{l,m}\int d\omega \Phi_1 Y_{l,m} e^{-i\omega t}.
\end{eqnarray}
Again the perturbations functions $(N, L, T, H_1, V, W, p_1, \rho_1, \Phi_1)$ depend only on the radial coordinate $r$, numbers $l$, $m$ and $\omega$.

In the following we will restrict to axial perturbations.

\subsection{Axial perturbations}

Introducing the axial perturbations into the field equations, one can obtain a set of differential and algebraic relations for the three perturbation functions. The minimal system of differential equations is given by

\begin{eqnarray}
\frac{d h_1}{dr} &=& i\omega \frac{F_1}{e^{2\nu} D_1} h_0 - \frac{F_2}{r D_1} h_1, \\
\frac{d h_0}{dr} &=& \frac{2}{r} h_0 + \left((l-1)(l+2)\frac{iF_3}{2\omega D_2} - i\omega \right) h_1,
\end{eqnarray}
and the relation
\begin{eqnarray}
w_2 &=& \frac{2i(\rho+p)}{\omega r^2}h_0.
\end{eqnarray}
The functions $F_1$, $F_2$, $F_3$, $D_1$ and $D_2$ are presented in Appendix \ref{ap_axial_fun}.

This system of equations can be rewritten as a second order ordinary differential equation for $h_1 = re^{\lambda-\nu}Z$:
\begin{equation}
\frac{d^2 Z}{dr^2} = A\frac{dZ}{dr} + (B\omega^2 + C)Z, \label{RW_EGBd}
\end{equation}
the generalized Regge-Wheeler equation for EGBd neutron stars \cite{PhysRev.108.1063}. The functions $A$, $B$ and $C$ are given in terms of the previous functions and derivatives, and can be found in Appendix \ref{ap_axial_fun}


In the limit when $\alpha=\beta=0$, the classical Regge-Wheeler equation for Einstein neutron stars is obtained. In the limit when $p=\rho=0$, the equations for axial perturbations of static EGBd black holes are reobtained \cite{2009PhRvD..79h4031P}.

This is essentially a Schr\"odinger equation where the eigenvalue $\omega=\omega_{\Re}+i\omega_{\Im}$ is a complex number. We consider 
perturbations 
exponentially damped with time. 

We require the perturbation to be regular at every point, in particular at the origin. This requires the Regge-Wheeler function to behave as
\begin{equation}
Z = Z_{l+1}r^{l+1} + O(r^{l+3}), \label{RW_org}
\end{equation}
for both the Einstein and EGBd cases.

We are interested in the resonant frequencies and damping times of gravitational waves coming out from the neutron star. Hence we require the wave to behave as a purely
outgoing wave at radial infinity. 
But in general a solution of the Regge-Wheeler equation will be
given by a superposition of incoming signal $Z^{in}$ and outgoing $Z^{out}$. Each component behaves like
\begin{equation}
\lim_{r\to\infty}Z^{in} \sim e^{i\omega r}, \lim_{r\to\infty}Z^{out} \sim e^{-i\omega r}. \label{asymp_out}
\end{equation}
Note that, while the real part of $\omega$ determines the oscillation frequency
of the wave, the imaginary part of the eigen-value determines the asymptotic
behavior of the quasi-normal mode: 
If we call $\tau=1/\omega_{\Im}$, and if we ignore the oscillating part of the function,
the ingoing and outgoing modes behave as
\begin{equation}
\lim_{r\to\infty}Z^{in} \sim e^{-r/\tau},  \lim_{r\to\infty}Z^{out} \sim e^{r/\tau},
\end{equation}
Outgoing quasi-normal modes are divergent
at radial infinity, while ingoing ones tend exponentially to zero as the
radius grows.

\subsection{Brief description of the numerical method} 

The divergent behavior of the purely outgoing wave is problematic from a numerical point of view. In
 principle, the purely outgoing quasi-normal mode 
 condition could be imposed at a very
 distant point, but because of the exponential decay of the ingoing signal at radial infinity,
 every small numerical error in the imposition of this behavior will be
 amplified as we approach the border of the star, resulting in a mixture of
 outgoing with ingoing waves.

To deal with this problem we use the method developed for Einstein theory in \cite{2013PhRvD..87j4042B,2014PhRvD..89d4006B}. Essentially, we integrate the full set of equations for the static configuration (\ref{TOV_EGBd_nu})-(\ref{dil_EGBd}) using a monotonic Hermite interpolation of several realistic tabulated EOS. We impose regularity conditions at the origin (\ref{bc_org}), and the correct asymptotic behavior at infinity (\ref{bc_inf}). Once we have the static configuration, we generate two independent solutions for the interior part of the Regge-Wheeler equation (\ref{RW_EGBd}), imposing the regularity condition (\ref{RW_org}). For the exterior part of the solution, we integrate the phase function, defined as $g = \frac{1}{Z}\frac{dZ}{dr}$, using the Exterior Complex Scaling method to guarantee that the solutions are outgoing waves \cite{2013PhRvD..87j4042B,2014PhRvD..89d4006B}. See Appendix \ref{app_ECS} for a more detailed explanation of the method.

For the numerical integration of the differential equations at each stage, we use Colsys \cite{colsys1979}, a package which implements a collocation method for boundary-value ordinary differential equations, equipped with an adaptive mesh selection
procedure. Typically the precision required to the solutions is better than $10^{-4}$, with meshes bigger than $100$ points.

Once we have these solutions for the perturbation, we have to study the matching conditions. In general, a quasi-normal mode will satisfy the usual junction conditions at the border of the star. This means that, in the absence of any surface energy density, the Regge-Wheeler function and its derivative must be continuous \cite{israel66}
.
These conditions are satisfied only if $\omega$ is a quasi-normal mode of the static configuration. 
As shown in Appendix \ref{app_ECS}, 
the junction conditions can be rewritten as a matrix, and continuity implies that the determinant of this matrix is zero. Once we fix the parameters of the theory ($\alpha$ and $\beta$), the static configuration (the mass $M$), and the angular $l$ number, this determinant is only a function of $\omega$. The quasi-normal modes are found by studying the zeros of this determinant.

\section{Results} \label{sec_results}

\begin{table}
\begin{center}
\begin{tabular}[b]{|c|c|l|}
\hline
\textbf{Model} & \textbf{EOS} & \textbf{Description} \\
\hline
        
Nuclear   &  SLy & Plain $npe\mu$ matter (potential method) \cite{Douchin2001}.  \\

 \cline{2-3}
matter  &  APR4 & Plain $npe\mu$ matter (variational method)  \cite{PhysRevC.58.1804}. \\
 
\hline
       &        & Nucleons and hyperons with a mixed phase \\
       & BHZBM  & (non-linear RMF method). Includes a first order \\
Hyperon&        & phase transition that softens the EOS \cite{Bednarek2012}. \\

 \cline{2-3}

matter & WCS1 & \lq\lq $\sigma \omega \rho \phi$ \rq\rq model for high densities. \\   
       & WCS2& The symmetric and anti-symmetric coupling ratios are \\ 
       &      & $\alpha_{\nu}=1$ and $\alpha_{\nu}=0.2$ respectively \cite{PhysRevC.85.065802,PhysRevC.90.019904}.  \\
\hline
       &     & Nuclear and quark matter, described by mixing\\
       & ALF4& the APR nuclear EOS with the CFL model of quark matter \cite{2005ApJ...629..969A}.\\

 \cline{2-3}

       &     & Hyperons and quark matter with a Maxwell phase transition. \\
Hybrid & BS4 & Hyper-nuclear density functional combined with effective \\
matter &     & NJL model of quantum chromodynamics. The parameters considered are \\
       &     & $G_V/G_S=0.6$, $\rho_{tr}/\rho_{0}=4$ \cite{Sed2012}.\\
 \cline{2-3}                        
       & WSPHS3      & Quark and hadronic matter with a Gibbs transition. MIT bag method \\
       & & mixed with NL3 RMF hadronic EOS with $B_{eff}^{1/4}=140 MeV$, $a_4=0.5$ \cite{2011ApJ...740L..14W}. \\
\hline

\end{tabular}
\caption{Equations of state considered in this paper. We have studied 2 nuclear matter EOS, 3 hyperonic matter EOS, and another 3 hybrid matter EOS.}
\label{tableEOS}
\end{center}
\end{table}

In Table \ref{tableEOS} we present the 8 realistic EOS considered in this paper. We consider two EOS for nuclear matter (SLy and APR4), three for hyperonic matter (WCS1-2 and BHZBM), and another three for hybrid quark-nuclear matter (ALF4, WSPHS3 and BS4).

\begin{figure}[h]
    \centering
    \includegraphics[width=0.45\textwidth,angle=-90]{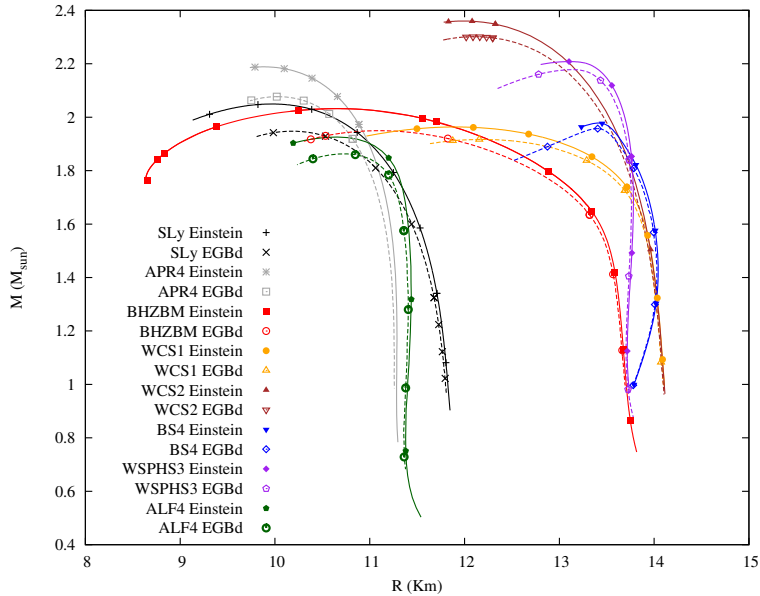}
    \caption{Mass vs Radius relation for the EOS considered, for pure Einstein and Einstein-Gauss-Bonnet with $\alpha_c$ and $\beta=1$.}
    \label{fig_M_R}
\end{figure}

In Figure \ref{fig_M_R} we plot the mass-radius diagram for these EOS, using pure Einstein theory, and 
EGBd theory with GB coupling $\alpha_c = 0.141 \alpha_Y$, where $\sqrt{|\alpha_Y|} = 1.9\times 10^{5}$ cm, the astrophysical constraint for the GB coupling as defined by Yagi.
This constraint was obtained in \cite{2012PhRvD..86h1504Y} studying the orbital decay rate of black holes in low-mass x-ray binaries. In addition, we consider the string theory value of the dilaton coupling $\beta = 1$. 

SLy, APR4 and ALF4 are the softest EOS considered, with radius between $10-12$ km. The other EOS are stiffer, with radius that go up to $14$ km. In the case of the WCS1 and BHZBM, the EOS become softer at higher densities. This causes the M-R curve of these two EOS to flatten near the maximum mass, decreasing the radius of these configurations.

Using the method previously discussed, we have studied axial quasi-normal modes of realistic neutron stars. In particular we have restricted to the $l=2$ fundamental curvature mode, and we will refer to it as wI mode \cite{lrr-1999-2,1742-6596-154-1-012039,PhysRevC.80.025801}. This is the standard family of pure space-time modes \cite{4727}. They were calculated for the first time for neutron stars in \cite{532}.

\subsection{Results for $l=2$ fundamental wI mode}

\begin{figure}[h]
    \centering
    \includegraphics[width=0.45\textwidth,angle=-90]{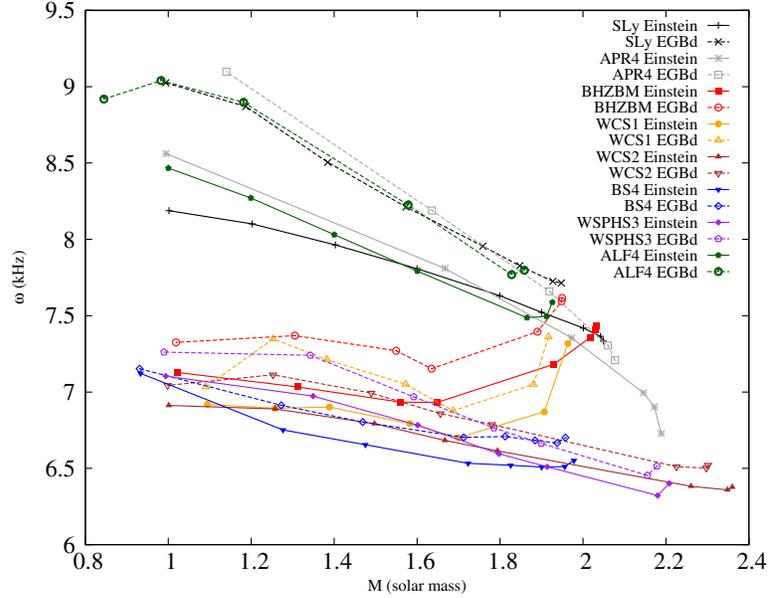}
    \caption{Frequency vs Mass for the $l=2$ fundamental wI mode for neutron stars between one solar mass and maximum mass. Note that the Gauss-Bonnet-dilaton, with $\alpha_c$ and $\beta=1$, increases the value of the frequency up to $10\%$.}
    \label{fig_M_freq_wI0}
\end{figure}

In Figure \ref{fig_M_freq_wI0} we show the frequency of the fundamental curvature modes versus the total mass. These frequencies are found in between $6$ kHz and $9$ kHz. For the stiffest EOS, the frequency is around 6.5-7.5 kHz, while for the softest EOS, the frequencies increase up to 9 kHz for the $1 M_{\odot}$ stars. 

The effect of the Gauss-Bonnet-dilaton term is always the same: the frequency increases with respect to a configuration of similar mass in Einstein theory. For the stiffest EOS, the effect is around $3-6 \%$, while for the softest it can go up to $10 \%$ for the low mass stars.

Note that for the BHZBM and the WCS1 EOS, the frequency increases for configurations between $1.6 M_{\odot}$ and $2 M_{\odot}$ masses. This effect is a manifestation of the softening of these EOS at higher densities. The softening causes the radius of the maximum mass configurations to decrease as the mass approaches the maximum value, increasing the value of the frequency. 

\begin{figure}[h]
    \centering
    \includegraphics[width=0.45\textwidth,angle=-90]{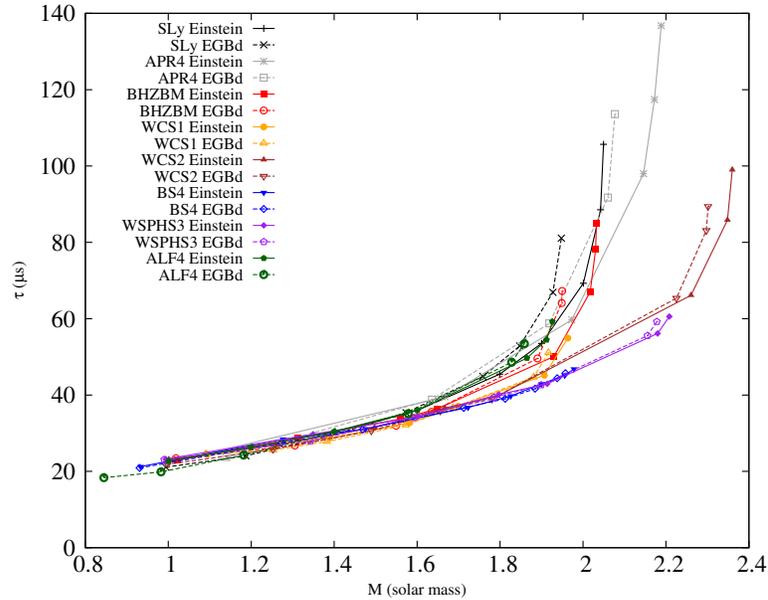}
    \caption{Damping time vs Mass for the $l=2$ fundamental wI mode for neutron stars between one solar mass and maximum mass. In this case note that the effect of the Gauss-Bonnet-dilaton term is negligible, except very close to the maximum masses.}
    \label{fig_M_tau_wI0}
\end{figure}

We present the same plot for the damping time $\tau$ in Figure \ref{fig_M_tau_wI0}. The damping time grows with the mass. The GBd term increases the damping time, but the effect is irrelevant for masses below $1.8 M_{\odot}$. It becomes more important close to the maximum mass, and specially, for EOS that become softer close to the maximum mass: for SLy, APR4, BHZBM and WC2, the increase in the damping time can be of $10 \%$.

\begin{figure}[h]
    \centering
    \includegraphics[width=0.45\textwidth,angle=-90]{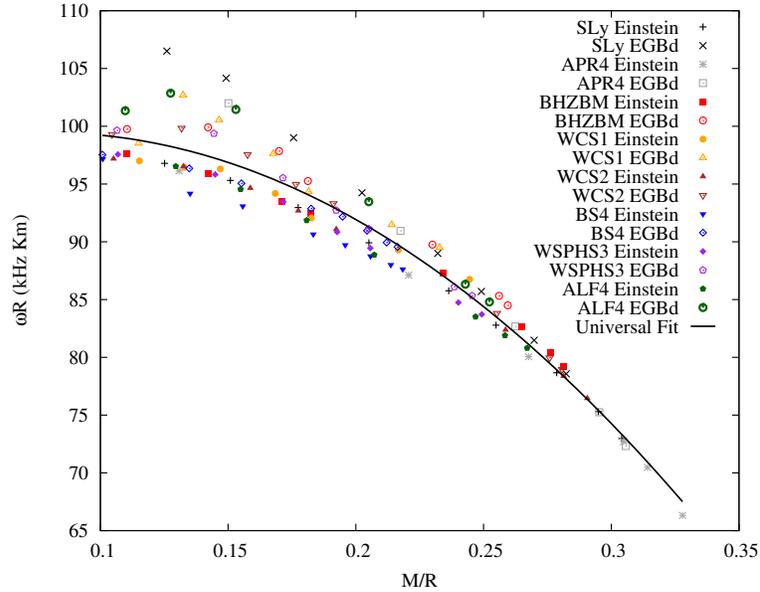}
    \caption{Frequency scaled to the radius vs compactness for the $l=2$ fundamental wI mode. A quadratic universal relation between the scaled frequency and the compactness holds approximately.}
    \label{fig_comp_Rfreq_wI0}
\end{figure}

\begin{figure}[h]
    \centering
    \includegraphics[width=0.45\textwidth,angle=-90]{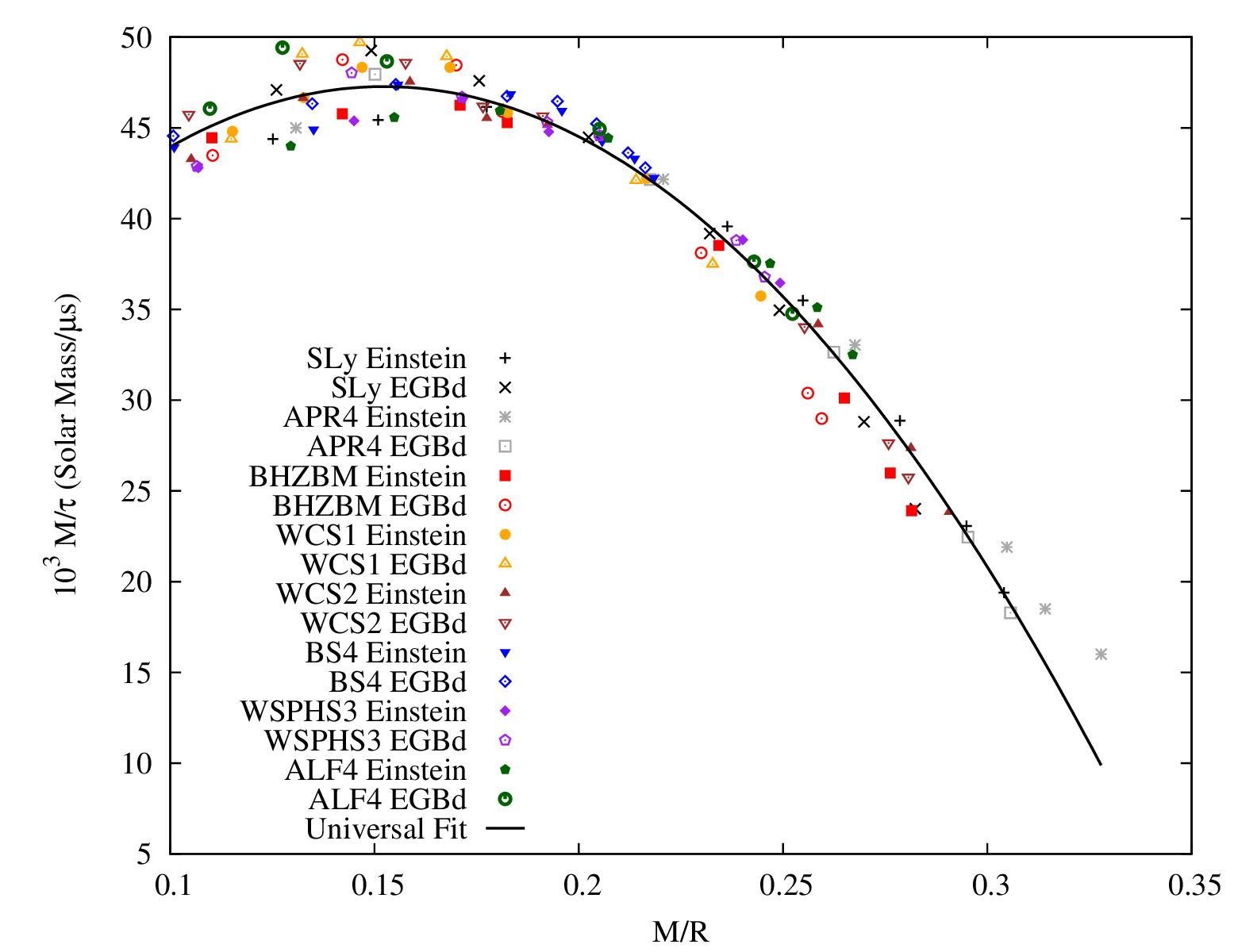}
    \caption{Damping scaled to the total mass vs compactness for the $l=2$ fundamental wI mode. A quadratic universal relation between the scaled damping time and the compactness holds approximately.}
    \label{fig_comp_Mtau_wI0}
\end{figure}

In the spirit of Refs. 
\cite{2001MNRAS.320..307K,2004PhRvD..70l4015B,PhysRevD.70.124015,PhysRevLett.77.4134,Andersson01101998}
, we study approximate universal relations of rescaled quantities. It is interesting to study if the introduction of the GBd term has some impact on the empirical relations valid for Einstein theory. In Figures \ref{fig_comp_Rfreq_wI0} and \ref{fig_comp_Mtau_wI0} we plot the scaled frequency and damping time versus the compactness. The frequency of the fundamental wI mode is given by the inverse of the radius, with some additional quadratic dependence with the compactness.  The inverse of the damping time is proportional to the inverse of the mass, with some additional quadratic dependence with the compactness. The phenomenological relations are
\begin{eqnarray}
\omega(khz) = \frac{1}{R(km)}\left[(-517.4\pm55.0)\left(\frac{M}{R}\right)^{2} + (82.2\pm22.6)\frac{M}{R} + (96.18\pm2.19)\right], \\
\label{empiricalfrec}
\frac{10^3}{\tau(\mu s)} = \frac{1}{M(M_{\odot})}\left[(-1213.02\pm39.14)\left(\frac{M}{R}\right)^{2} + (369.517\pm16.09)\frac{M}{R} + (19.132\pm1.557)\right].
\label{empiricalomgtau}
\end{eqnarray}    
They hold approximately for every EOS, even in the presence of the GBd term.

\begin{figure} [h]
    \centering
    \includegraphics[width=0.6\textwidth,angle=-90]{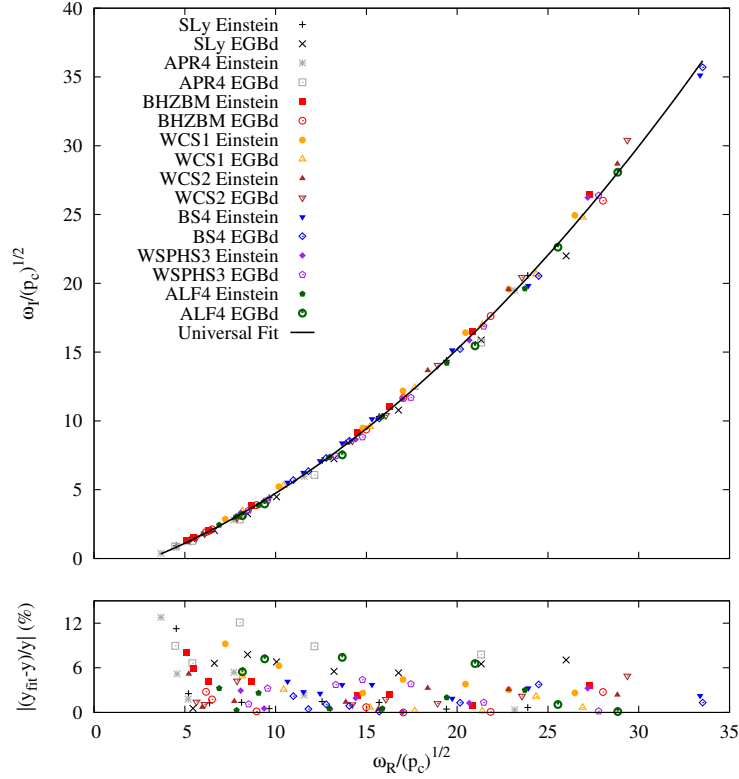}
    \caption{$\frac{\omega_I}{\sqrt{p_c}}$ vs $\frac{\omega_R}{\sqrt{p_c}}$ for the $l=2$ fundamental wI mode. A quadratic universal relation holds approximately.}
    \label{fig_omR_omI_wI0_error}
\end{figure}

In \cite{2013PhRvD..87j4042B,2014PhRvD..89d4006B} it was shown that another interesting universal relation could be obtained if the central pressure is used as scaling factor, instead of global parameters like the radius or the total mass. We show the effect of the GBd term in this alternative scaling in Figure \ref{fig_omR_omI_wI0_error}. A quadratic relation between the scaled imaginary part and the scaled real part is obtained:

\begin{eqnarray}
\frac{\omega_{\Im}}{\sqrt{p_c}}=(-1.45 \pm 0.16) +  (0.405 \pm 0.022)\frac{\omega_{\Re}}{\sqrt{p_c}} + 
(0.0214 \pm 0.0006)\left(\frac{\omega_{\Re}}{\sqrt{p_c}}\right)^{2}. \label{pc_scaled_modes}
\end{eqnarray}

This universal relation is also approximately independent of the EOS and valid for both Einstein and EGBd gravity.

Hence we conclude that, since the effect of the EGBd theory on the phenomenological relations is very small, these empirical formulas could also be used in the study of EGBd gravity. On the other hand, frequencies in EGBd gravity are sensitively bigger than the frequency of the Einstein gravity modes. This could have some observational implications, as we consider in the next section.

\subsection{Possible application of the universal relation to a detection}

These universal relations can be used to estimate some parameters of the star, and in some cases further constrain the valid equations of state or GBd coupling parameters. Let us consider two examples.

\begin{figure}
    \centering
    \includegraphics[width=0.45\textwidth,angle=-90]{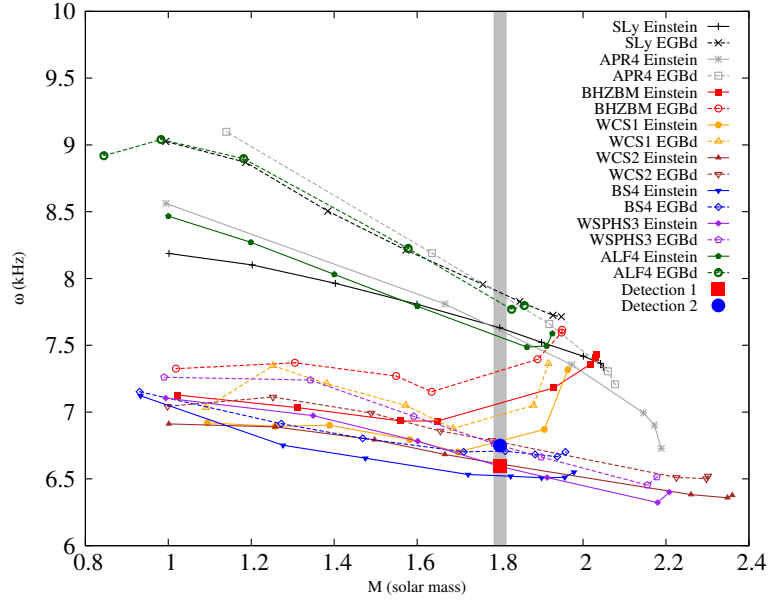}
    \caption{Frequency vs Mass for the $l=2$ fundamental wI mode for neutron stars between one solar mass and maximum mass. To possible detections of gravitational waves coming from a neutron star of 1.8 solar mass are considered: Detection 1 with frequency of $6.6$ kHz, damping time of $40$ $\mu$s; and detection 2 with $6.75$ kHz and same damping time.}
    \label{fig_M_freq_wI0_detection}
\end{figure}

\begin{figure}
    \centering
    \includegraphics[width=0.45\textwidth,angle=-90]{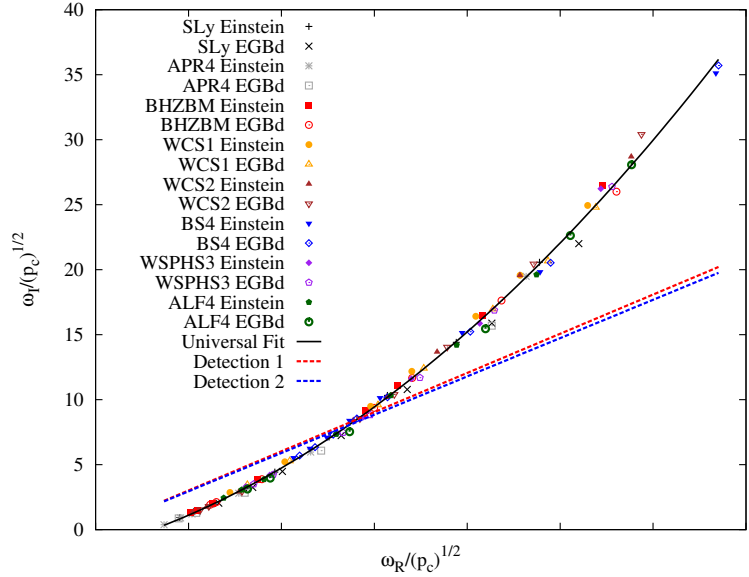}
    \caption{Universal relation between $\omega_R$ and $\omega_I$ scaled to the central pressure, for the $l=2$ fundamental wI mode. Each one of the detections defines a line in the plot, that intersects the universal relation at a different point, determining an estimated central pressure. For detection 1, the central pressure is $1.17\times10^{35} dyn/cm^2$. For detection 2, the central pressure is $1.31\times10^{35} dyn/cm^2$.}
    \label{fig_omR_omI_wI0_detection}
\end{figure}

\begin{figure}
    \centering
    \includegraphics[width=0.45\textwidth,angle=-90]{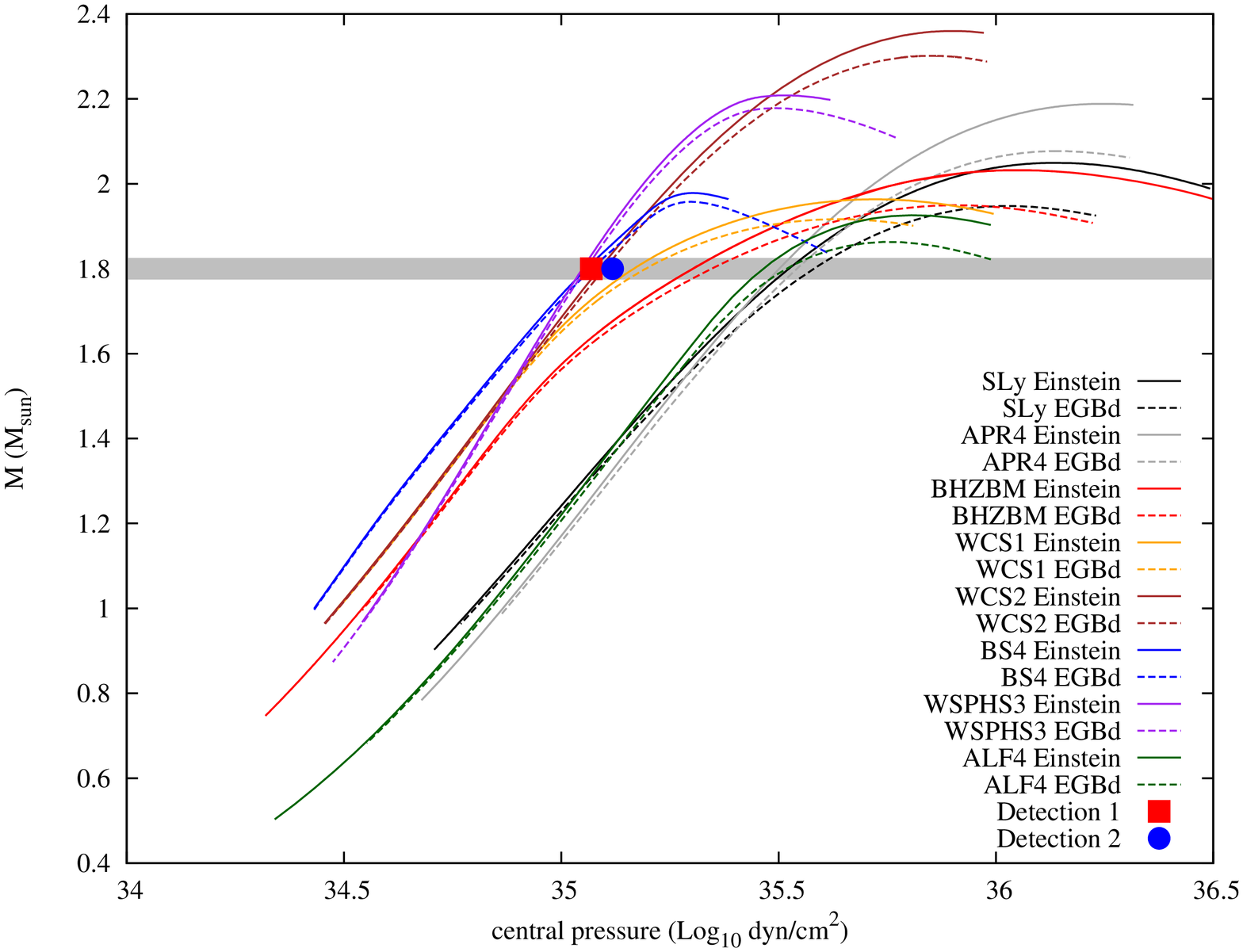}
    \caption{Mass vs central pressure. As explained in the text, the estimated central pressure could allow us in some cases to constrain the EOS and estimate the Gauss-Bonnet-dilaton coupling.}
    \label{fig_M_pc_detection}
\end{figure}

Consider a possible gravitational wave detected from a neutron star with $1.8 M_{\odot}$. In detection 1, the frequency of the fundamental wI mode is determined to be $6.6$ kHz with a damping time of $40$ $\mu s$. Assuming a $1 \%$ error, we plot this detection as a red square in Figure \ref{fig_M_freq_wI0_detection}. This detection would constrain the possible matter constitution to the EOS WCS2, WSPHS3 in Einstein theory, or EOS BS4 with a non-vanishing GBd coupling. This detection would draw a line in the central pressure scaled plot (see red line in Figure \ref{fig_omR_omI_wI0_detection}). The crossing point with the universal relation (\ref{pc_scaled_modes}) determines a central pressure ($p_c=1.17\times10^{35} dyn/cm^{2}$). With this estimation we can mark the configuration in Figure \ref{fig_M_pc_detection}, where we plot the total mass versus the central pressure. In this case, the three EOS are still compatible with the central pressure estimation, and the phenomenological relation does not give us more information about the theory or matter constitution.

Now consider an alternative scenario, detection 2, where the frequency of the fundamental wI mode is determined to be $6.75$ kHz with a similar damping time of $40$ $\mu s$. This is marked in Figure \ref{fig_M_freq_wI0_detection} as the blue dot. In this case the detection would be compatible with Einstein theory with EOS WCS1, or EGBd theory with EOS WCS2, BS4 or WSPHS3. With this information we can use the phenomenological relation (\ref{pc_scaled_modes}) to estimate the central pressure. We draw a blue line in Figure \ref{fig_omR_omI_wI0_detection}, and the estimated central pressure is $p_c=1.31\times10^{35} dyn/cm^{2}$ in this case. If we mark this configuration in Figure \ref{fig_M_pc_detection} (blue dot), we can see that the only EOS compatible with this estimated central pressure are reduced to two cases: Einstein theory with WCS1 or EGBd theory with WCS2 EOS. In this case, assuming that the phenomenological relation holds and the precision of the detection is good enough, we are able to further constrain the equation of state and coupling parameters of the theory. 

The two arbitrary scenarios presented show us how the phenomenological relation (\ref{pc_scaled_modes}) could be used in a future true detection. Although in some cases the phenomenological relation does not give us more information about the structure of the star, as shown in the example we called detection 1, we have shown that there are some favorable cases, as shown in the example detection 2, where these phenomenological relations could be used to further constrain the composition or the theory describing the neutron star.
 
\section{Conclusions} \label{sec_conclusions}

We have considered axial quasi-normal modes of realistic neutron stars in Einstein-Gauss-Bonnet-dilaton gravity. We have focused the study on the fundamental space-time curvature mode (wI mode) for 8 different realistic equations of state. 

The general equations for axial quasi-normal modes of neutron stars have been derived, and we have developed a method to obtain the spectrum of QNM using realistic tabulated EOS. The EOS considered include plain nuclear matter, hyperon matter, and hybrid matter.

We have analyzed the effect of the GBd terms on the QNM by comparing with similar modes of configurations in pure Einstein gravity. We have found that the GBd term increases the frequency of the modes, which can go up to $10\%$ for a large Gauss-Bonnet coupling, while the effect on the damping time is typically smaller.

Interestingly, we have found that the universal relations valid in Einstein theory for curvature modes, still hold in EGBd gravity, for relations scaling to global parameters as mass and radius, or the central pressure. 

We have shown that in some cases, these universal relations, together with the fact that the frequency of the wI mode is increased in a EGBd star compared with a star of similar mass in Einstein gravity, could be used to constrain the equations of state and possible GBd coupling of the theory.   

Although in this paper we have focused on the fundamental wI mode, a possible next step is the study of interface modes and higher excitations of the curvature modes, although more precision and calculation time will be required.

The next step is the study of polar QNM for realistic neutron stars in EGBd theory. For this, the calculation of the equations for polar perturbations is necessary. Polar modes in general include the families of pressure modes, the fundamental mode, as well as another branch of curvature modes. Since polar modes allow for scalar perturbations, they are more sensitive to the matter composition of the neutron star. But note that in EGBd theory, polar perturbations would allow for oscillations of the dilaton field. It will be interesting to study the possible effect of the dilaton on the polar modes, and to see whether new families of modes not present in Einstein gravity could be present in EGBd theory.

\section{Acknowledgments}

JLBS, JK and SM gratefully acknowledge support by the DFG
Research Training Group 1620 “Models of Gravity”, as well as support from FP7,
Marie Curie Actions, People, International Research Staff Exchange Scheme
(IRSES-606096). 
JLBS, LMGR, JK and FNL gratefully acknowledge support from MINECO, under
research Project No. FIS2011-28013.

\appendix

\section{Equations for the axial perturbations} \label{ap_axial_fun}

The equations describing the axial perturbations are

\begin{eqnarray}
\frac{d h_1}{dr} &=& i\omega \frac{F_1}{e^{2\nu} D_1} h_0 - \frac{F_2}{r D_1} h_1, \\
\frac{d h_0}{dr} &=& \frac{2}{r} h_0 + \left((l-1)(l+2)\frac{iF_3}{2\omega D_2} - i\omega \right) h_1, \\
w_2 &=& \frac{2i(\rho+p)}{\omega r^2}h_0.
\end{eqnarray}
where the functions $F_1$, $F_2$, $F_3$, $D_1$ and $D_2$ are given by the static solution, and can be written as
\begin{eqnarray}
F_1 &=& 4\alpha\beta r (r-2m)\left[\beta\left({\frac{d\Phi}{dr}}\right)^{2}-\frac{d^2\Phi}{dr^2}\right]+4\alpha\beta\frac{d\Phi}{dr}\left(r\frac{dm}{dr}-m\right) + e^{\beta\Phi}r^2
, \\
F_2 &=& 4\alpha\beta\left[4mr(m-r)+r^3\right]\left[\frac{d^2\Phi}{dr^2}\frac{d\nu}{dr}+\frac{d^2\nu}{dr^2}\frac{d\Phi}{dr}+
\frac{d\Phi}{dr}\left(\frac{d\nu}{dr}\right)^2 -\beta\frac{d\nu}{dr}\left(\frac{d\Phi}{dr}\right)^2 \right] \nonumber \\ &+&
12\alpha\beta\frac{d\Phi}{dr}\frac{d\nu}{dr}\left[m(r-2m)+(2mr-r^2)\frac{dm}{dr}\right] 
+ re^{\beta\Phi}\left[r\frac{d\nu}{dr}(2m-r)+r\frac{dm}{dr}-m\right]
,\\
F_3 &=& 
e^{\beta\Phi+2\nu}\left[
2r^2e^{\beta\Phi}
- \alpha\beta r^2 \left({\frac{d\Phi}{dr}}\right)^{3}\left(r-2m\right) 
-8\frac{d\Phi}{dr}\alpha\beta\left((r-2m)
+ 4\pi p r^3\right)
\right]
,\\
D_1 &=&   \left(2m-r \right)  \left( 
4\alpha\beta\frac{d\Phi}{dr}\frac{d\nu}{dr}(2m-r)+e^{\beta\Phi}r
\right),\\
D_2 &=& 
\left[4\alpha\beta\frac{d\Phi}{dr}(3m-r)+e^{\beta\Phi}r^2\right]\left[4\alpha\beta\frac{d\Phi}{dr}(2m-r)+e^{\beta\Phi}r^2\right]
.
\end{eqnarray}

The generalized Regge-Wheeler equation for EGBd neutron stars is obtained substituting $h_1 = re^{\lambda-\nu}Z$:
\begin{equation}
\frac{d^2 Z}{dr^2} = A\frac{dZ}{dr} + (B\omega^2 + C)Z, \label{RW_EGBd_app}
\end{equation}
where the functions $A$, $B$ and $C$ are given in terms of the previous functions and derivatives and can be written as
\begin{eqnarray}
A &=& \frac{d}{dr}\ln{\left(\frac{F_1}{D_1}\right)}-\frac{F_2}{rD_1}-2\frac{d\lambda}{dr}, \\
B &=& e^{-2\nu}\frac{F_1}{D_1}, \\
C &=& \frac{2}{r^2} - e^{-2\nu}\frac{l(l+1)}{2}\frac{F_1 F_3}{D_1 D_2} + \frac{d^2\nu}{dr^2}- \frac{d^2\lambda}{dr^2} + \left(\frac{d\nu}{dr}\right)^2- \left(\frac{d\lambda}{dr}\right)^2
+ \frac{1}{r}\frac{d}{dr}\ln{\left(\frac{F_1}{D_1}\right)} \nonumber \\ &+& \frac{F_2}{r F_1 D_1} \frac{dF_1}{dr} - \frac{1}{r D_1} \frac{dF_2}{dr}  +
e^{-2\nu} \frac{F_1 F_3}{D_1 D_2} + 2 \frac{F_2}{r^2 D_1}
+ \left[\frac{d}{dr}\ln{\left(\frac{F_1}{D_1}\right)}- \frac{F_2}{r D_1}\right]\frac{d\lambda}{dr} 
\nonumber \\ &-& \left[ \frac{d}{dr}\ln{\left(\frac{F_1}{D_1}\right)} + \frac{F_2}{r D_1} + \frac{2}{r}\right]\frac{d\nu}{dr}.
\label{eq:ABC}
\end{eqnarray}

\section{Exterior Complex Scaling and junction conditions}\label{app_ECS} 

%
%

The first step of our method is the integration of the interior part of the Regge-Wheeler solution. We generate two linearly independent solutions of the Regge-Wheeler equation, $Z^{(s1)}_{in}$ and $Z^{(s2)}_{in}$, with two different complex values at the border of the star $Z^{(s1)}_{in}(R) = Z^{(s1)}_{in,\Re}(R) + iZ^{(s1)}_{in,\Re}(R)$ and $Z^{(s2)}(R) = Z^{(s2)}_{in,\Re}(R) + iZ^{(s2)}_{in,\Re}(R)$. Both solutions satisfy the regularity condition at the origin.

The next step is to calculate the exterior part of the solution
requiring the outgoing wave behavior. We choose to integrate the phase function defined as $g = \frac{1}{Z}\frac{dZ}{dr}$, rewriting the Regge-Wheeler equation as
\begin{equation}
\frac{dg}{dr} = -g^2 + Ag + B\omega^2 + C.
\end{equation}
At radial infinity the phase function of a mixture of ingoing and outgoing wave takes the value $g(\infty)=-i\omega$ 
. To remove the ingoing wave part, we use the standard Exterior Complex Scaling method. We change to a complex radial coordinate $r = R + ye^{ia_0}$, where $a_0$ satisfies
 \begin{eqnarray}
 \omega_{\Re}\sin a_0 + \omega_{\Im}\cos a_0 < 0.
 \end{eqnarray}
Now by requiring the boundary condition $g(\infty)=-i\omega$, we grant that the exterior part of the solution has no ingoing wave contamination.

Once we have these solutions for the perturbation, we have to study the matching conditions. In general, a quasi-normal mode will satisfy the usual junction conditions at the border of the star. In the absence of a surface energy density at the border of the star, the junction conditions are given by the continuity of the first and second fundamental forms defined on the surface of the star  \cite{israel66}.
This implies that the Regge-Wheeler function and its derivative must be continuous at the border:
\begin{eqnarray}
Z_{out}(R) = Z_{in}(R), \label{junction_eqs1}\\
\frac{d}{dr}Z_{out}(R) = \frac{d}{dr}Z_{in}(R). \label{junction_eqs2}
\end{eqnarray}
These conditions are satisfied only if $\omega$ is a quasi-normal mode of the static configuration. Since we have two independent solutions inside the star, and the phase outside, we can impose the continuity of the function for both solutions. This means we have two independent values of the derivative of the Regge-Wheeler function outside the star:
\begin{eqnarray}
\frac{d}{dr}Z^{(s1)}_{out}(R) = Z^{(s1)}_{in}(R) g(R), \\
\frac{d}{dr}Z^{(s2)}_{out}(R) = Z^{(s2)}_{in}(R) g(R).
\end{eqnarray}
For some values of the parameters of the solution, we will have that a linear combination of the two solutions $s1$ and $s2$ satisfy the continuity requirement. Hence, we can rewrite the junction conditions (\ref{junction_eqs1}) and (\ref{junction_eqs2}) as
\begin{eqnarray}
K_1 Z^{(s1)}_{in}(R) + K_2 Z^{(s2)}_{in}(R) = K_1 Z^{(s1)}_{out}(R) + K_2 Z^{(s2)}_{out}(R), \label{junction_eqs1_lin} \\
K_1 \frac{d}{dr}Z^{(s1)}_{in}(R) + K_2 \frac{d}{dr}Z^{(s2)}_{in}(R) = K_1 \frac{d}{dr}Z^{(s1)}_{out}(R) + K_2 \frac{d}{dr}Z^{(s2)}_{out}(R). \label{junction_eqs2_lin}
\end{eqnarray}
The first condition (\ref{junction_eqs1_lin}) is trivially satisfied since each particular solution is continuous. But the continuity of the first derivative (\ref{junction_eqs2_lin}) can be rewritten in matrix form:
\begin{equation}
\hat{M} 
\left[
\begin{array}{c}
K_1 \\
K_2
\end{array}
\right]
=
\left[
\begin{array}{c c}
\frac{d}{dr}Z^{(s1)}_{out,\Re}(R)-\frac{d}{dr}Z^{(s1)}_{in,\Re}(R) & \frac{d}{dr}Z^{(s2)}_{out,\Re}(R)-\frac{d}{dr}Z^{(s2)}_{in,\Re}(R) \\
\frac{d}{dr}Z^{(s1)}_{out,\Im}(R)-\frac{d}{dr}Z^{(s1)}_{in,\Im}(R) & \frac{d}{dr}Z^{(s2)}_{out,\Im}(R)-\frac{d}{dr}Z^{(s2)}_{in,\Im}(R)
\end{array}
\right]
\left[
\begin{array}{c}
K_1 \\
K_2
\end{array}
\right]
=0.
\end{equation}

The quasi-normal modes are found when the determinant of the $\hat{M}$ matrix is zero. This determinant is only a function of $\omega$, once we have fixed the theory ($\alpha$ and $\beta$), the static configuration (the mass $M$), and the $l$ number.


\bibliography{bib_NS_EGBd_axial_QNM}

\begin{thebibliography}{44}
\expandafter\ifx\csname natexlab\endcsname\relax\def\natexlab#1{#1}\fi
\expandafter\ifx\csname bibnamefont\endcsname\relax
  \def\bibnamefont#1{#1}\fi
\expandafter\ifx\csname bibfnamefont\endcsname\relax
  \def\bibfnamefont#1{#1}\fi
\expandafter\ifx\csname citenamefont\endcsname\relax
  \def\citenamefont#1{#1}\fi
\expandafter\ifx\csname url\endcsname\relax
  \def\url#1{\texttt{#1}}\fi
\expandafter\ifx\csname urlprefix\endcsname\relax\def\urlprefix{URL }\fi
\providecommand{\bibinfo}[2]{#2}
\providecommand{\eprint}[2][]{\url{#2}}

\bibitem[{\citenamefont{Will}(2014)}]{clifford:LRR2014}
\bibinfo{author}{\bibfnamefont{C.~M.} \bibnamefont{Will}},
  \bibinfo{journal}{Living Rev.Rel.} \textbf{\bibinfo{volume}{17}},
  \bibinfo{pages}{4} (\bibinfo{year}{2014}), \eprint{1403.7377}.

\bibitem[{\citenamefont{Haensel et~al.}(2006)\citenamefont{Haensel, Potekhin,
  and Yakovlev}}]{haensel2006neutron}
\bibinfo{author}{\bibfnamefont{P.}~\bibnamefont{Haensel}},
  \bibinfo{author}{\bibfnamefont{A.}~\bibnamefont{Potekhin}}, \bibnamefont{and}
  \bibinfo{author}{\bibfnamefont{D.}~\bibnamefont{Yakovlev}},
  \emph{\bibinfo{title}{Neutron Stars 1: Equation of State and Structure}},
  Astrophysics and Space Science Library (\bibinfo{publisher}{Springer},
  \bibinfo{year}{2006}), ISBN \bibinfo{isbn}{9780387335438},
  \urlprefix\url{http://books.google.de/books?id=iIrj9nfHnesC}.

\bibitem[{\citenamefont{{Demorest} et~al.}(2010)\citenamefont{{Demorest},
  {Pennucci}, {Ransom}, {Roberts}, and {Hessels}}}]{2010Natur.467.1081D}
\bibinfo{author}{\bibfnamefont{P.~B.} \bibnamefont{{Demorest}}},
  \bibinfo{author}{\bibfnamefont{T.}~\bibnamefont{{Pennucci}}},
  \bibinfo{author}{\bibfnamefont{S.~M.} \bibnamefont{{Ransom}}},
  \bibinfo{author}{\bibfnamefont{M.~S.~E.} \bibnamefont{{Roberts}}},
  \bibnamefont{and} \bibinfo{author}{\bibfnamefont{J.~W.~T.}
  \bibnamefont{{Hessels}}}, \bibinfo{journal}{\nat}
  \textbf{\bibinfo{volume}{467}}, \bibinfo{pages}{1081} (\bibinfo{year}{2010}),
  \eprint{1010.5788}.

\bibitem[{\citenamefont{Antoniadis et~al.}(2013)\citenamefont{Antoniadis,
  Freire, Wex, Tauris, Lynch, van Kerkwijk, Kramer, Bassa, Dhillon, Driebe
  et~al.}}]{Antoniadis26042013}
\bibinfo{author}{\bibfnamefont{J.}~\bibnamefont{Antoniadis}},
  \bibinfo{author}{\bibfnamefont{P.~C.~C.} \bibnamefont{Freire}},
  \bibinfo{author}{\bibfnamefont{N.}~\bibnamefont{Wex}},
  \bibinfo{author}{\bibfnamefont{T.~M.} \bibnamefont{Tauris}},
  \bibinfo{author}{\bibfnamefont{R.~S.} \bibnamefont{Lynch}},
  \bibinfo{author}{\bibfnamefont{M.~H.} \bibnamefont{van Kerkwijk}},
  \bibinfo{author}{\bibfnamefont{M.}~\bibnamefont{Kramer}},
  \bibinfo{author}{\bibfnamefont{C.}~\bibnamefont{Bassa}},
  \bibinfo{author}{\bibfnamefont{V.~S.} \bibnamefont{Dhillon}},
  \bibinfo{author}{\bibfnamefont{T.}~\bibnamefont{Driebe}},
  \bibnamefont{et~al.}, \bibinfo{journal}{Science}
  \textbf{\bibinfo{volume}{340}} (\bibinfo{year}{2013}),
  \eprint{http://www.sciencemag.org/content/340/6131/1233232.full.pdf},
  \urlprefix\url{http://www.sciencemag.org/content/340/6131/1233232.abstract}.

\bibitem[{\citenamefont{Pitkin et~al.}(2011)\citenamefont{Pitkin, Reid, Rowan,
  and Hough}}]{Pitkin:LRR2011}
\bibinfo{author}{\bibfnamefont{M.}~\bibnamefont{Pitkin}},
  \bibinfo{author}{\bibfnamefont{S.}~\bibnamefont{Reid}},
  \bibinfo{author}{\bibfnamefont{S.}~\bibnamefont{Rowan}}, \bibnamefont{and}
  \bibinfo{author}{\bibfnamefont{J.}~\bibnamefont{Hough}},
  \bibinfo{journal}{Living Rev.Rel.} \textbf{\bibinfo{volume}{14}},
  \bibinfo{pages}{5} (\bibinfo{year}{2011}), \eprint{1102.3355}.

\bibitem[{\citenamefont{Kokkotas and Schmidt}(1999)}]{lrr-1999-2}
\bibinfo{author}{\bibfnamefont{K.~D.} \bibnamefont{Kokkotas}} \bibnamefont{and}
  \bibinfo{author}{\bibfnamefont{B.~G.} \bibnamefont{Schmidt}},
  \bibinfo{journal}{Living Reviews in Relativity} \textbf{\bibinfo{volume}{2}}
  (\bibinfo{year}{1999}),
  \urlprefix\url{http://www.livingreviews.org/lrr-1999-2}.

\bibitem[{\citenamefont{Nollert}(1999)}]{0264-9381-16-12-201}
\bibinfo{author}{\bibfnamefont{H.-P.} \bibnamefont{Nollert}},
  \bibinfo{journal}{Classical and Quantum Gravity}
  \textbf{\bibinfo{volume}{16}}, \bibinfo{pages}{R159} (\bibinfo{year}{1999}),
  \urlprefix\url{http://stacks.iop.org/0264-9381/16/i=12/a=201}.

\bibitem[{\citenamefont{{Ferrari} and {Gualtieri}}(2008)}]{2008GReGr..40..945F}
\bibinfo{author}{\bibfnamefont{V.}~\bibnamefont{{Ferrari}}} \bibnamefont{and}
  \bibinfo{author}{\bibfnamefont{L.}~\bibnamefont{{Gualtieri}}},
  \bibinfo{journal}{General Relativity and Gravitation}
  \textbf{\bibinfo{volume}{40}}, \bibinfo{pages}{945} (\bibinfo{year}{2008}),
  \eprint{0709.0657}.

\bibitem[{\citenamefont{Moura and Schiappa}(2007)}]{0264-9381-24-2-006}
\bibinfo{author}{\bibfnamefont{F.}~\bibnamefont{Moura}} \bibnamefont{and}
  \bibinfo{author}{\bibfnamefont{R.}~\bibnamefont{Schiappa}},
  \bibinfo{journal}{Classical and Quantum Gravity}
  \textbf{\bibinfo{volume}{24}}, \bibinfo{pages}{361} (\bibinfo{year}{2007}),
  \urlprefix\url{http://stacks.iop.org/0264-9381/24/i=2/a=006}.

\bibitem[{\citenamefont{{Kanti} et~al.}(2011)\citenamefont{{Kanti}, {Kleihaus},
  and {Kunz}}}]{2011PhRvL.107A1101K}
\bibinfo{author}{\bibfnamefont{P.}~\bibnamefont{{Kanti}}},
  \bibinfo{author}{\bibfnamefont{B.}~\bibnamefont{{Kleihaus}}},
  \bibnamefont{and} \bibinfo{author}{\bibfnamefont{J.}~\bibnamefont{{Kunz}}},
  \bibinfo{journal}{Physical Review Letters} \textbf{\bibinfo{volume}{107}},
  \bibinfo{eid}{271101} (\bibinfo{year}{2011}), \eprint{1108.3003}.

\bibitem[{\citenamefont{{Kanti} et~al.}(2012)\citenamefont{{Kanti}, {Kleihaus},
  and {Kunz}}}]{2012PhRvD..85d4007K}
\bibinfo{author}{\bibfnamefont{P.}~\bibnamefont{{Kanti}}},
  \bibinfo{author}{\bibfnamefont{B.}~\bibnamefont{{Kleihaus}}},
  \bibnamefont{and} \bibinfo{author}{\bibfnamefont{J.}~\bibnamefont{{Kunz}}},
  \bibinfo{journal}{\prd} \textbf{\bibinfo{volume}{85}}, \bibinfo{eid}{044007}
  (\bibinfo{year}{2012}), \eprint{1111.4049}.

\bibitem[{\citenamefont{{Berti} et~al.}(2015)\citenamefont{{Berti}, {Barausse},
  {Cardoso}, {Gualtieri}, {Pani}, {Sperhake}, {Stein}, {Wex}, {Yagi}, {Baker}
  et~al.}}]{2015arXiv150107274B}
\bibinfo{author}{\bibfnamefont{E.}~\bibnamefont{{Berti}}},
  \bibinfo{author}{\bibfnamefont{E.}~\bibnamefont{{Barausse}}},
  \bibinfo{author}{\bibfnamefont{V.}~\bibnamefont{{Cardoso}}},
  \bibinfo{author}{\bibfnamefont{L.}~\bibnamefont{{Gualtieri}}},
  \bibinfo{author}{\bibfnamefont{P.}~\bibnamefont{{Pani}}},
  \bibinfo{author}{\bibfnamefont{U.}~\bibnamefont{{Sperhake}}},
  \bibinfo{author}{\bibfnamefont{L.~C.} \bibnamefont{{Stein}}},
  \bibinfo{author}{\bibfnamefont{N.}~\bibnamefont{{Wex}}},
  \bibinfo{author}{\bibfnamefont{K.}~\bibnamefont{{Yagi}}},
  \bibinfo{author}{\bibfnamefont{T.}~\bibnamefont{{Baker}}},
  \bibnamefont{et~al.}, \bibinfo{journal}{ArXiv e-prints}
  (\bibinfo{year}{2015}), \eprint{1501.07274}.

\bibitem[{\citenamefont{{Kanti} et~al.}(1996)\citenamefont{{Kanti},
  {Mavromatos}, {Rizos}, {Tamvakis}, and {Winstanley}}}]{1996PhRvD..54.5049K}
\bibinfo{author}{\bibfnamefont{P.}~\bibnamefont{{Kanti}}},
  \bibinfo{author}{\bibfnamefont{N.~E.} \bibnamefont{{Mavromatos}}},
  \bibinfo{author}{\bibfnamefont{J.}~\bibnamefont{{Rizos}}},
  \bibinfo{author}{\bibfnamefont{K.}~\bibnamefont{{Tamvakis}}},
  \bibnamefont{and}
  \bibinfo{author}{\bibfnamefont{E.}~\bibnamefont{{Winstanley}}},
  \bibinfo{journal}{\prd} \textbf{\bibinfo{volume}{54}}, \bibinfo{pages}{5049}
  (\bibinfo{year}{1996}), \eprint{hep-th/9511071}.

\bibitem[{\citenamefont{{Pani} and {Cardoso}}(2009)}]{2009PhRvD..79h4031P}
\bibinfo{author}{\bibfnamefont{P.}~\bibnamefont{{Pani}}} \bibnamefont{and}
  \bibinfo{author}{\bibfnamefont{V.}~\bibnamefont{{Cardoso}}},
  \bibinfo{journal}{\prd} \textbf{\bibinfo{volume}{79}}, \bibinfo{eid}{084031}
  (\bibinfo{year}{2009}), \eprint{0902.1569}.

\bibitem[{\citenamefont{{Pani} et~al.}(2011{\natexlab{a}})\citenamefont{{Pani},
  {Macedo}, {Crispino}, and {Cardoso}}}]{2011PhRvD..84h7501P}
\bibinfo{author}{\bibfnamefont{P.}~\bibnamefont{{Pani}}},
  \bibinfo{author}{\bibfnamefont{C.~F.~B.} \bibnamefont{{Macedo}}},
  \bibinfo{author}{\bibfnamefont{L.~C.~B.} \bibnamefont{{Crispino}}},
  \bibnamefont{and}
  \bibinfo{author}{\bibfnamefont{V.}~\bibnamefont{{Cardoso}}},
  \bibinfo{journal}{\prd} \textbf{\bibinfo{volume}{84}}, \bibinfo{eid}{087501}
  (\bibinfo{year}{2011}{\natexlab{a}}), \eprint{1109.3996}.

\bibitem[{\citenamefont{{Kleihaus} et~al.}(2011)\citenamefont{{Kleihaus},
  {Kunz}, and {Radu}}}]{2011PhRvL.106o1104K}
\bibinfo{author}{\bibfnamefont{B.}~\bibnamefont{{Kleihaus}}},
  \bibinfo{author}{\bibfnamefont{J.}~\bibnamefont{{Kunz}}}, \bibnamefont{and}
  \bibinfo{author}{\bibfnamefont{E.}~\bibnamefont{{Radu}}},
  \bibinfo{journal}{Physical Review Letters} \textbf{\bibinfo{volume}{106}},
  \bibinfo{eid}{151104} (\bibinfo{year}{2011}), \eprint{1101.2868}.

\bibitem[{\citenamefont{{Kleihaus} et~al.}(2014)\citenamefont{{Kleihaus},
  {Kunz}, and {Mojica}}}]{2014PhRvD..90f1501K}
\bibinfo{author}{\bibfnamefont{B.}~\bibnamefont{{Kleihaus}}},
  \bibinfo{author}{\bibfnamefont{J.}~\bibnamefont{{Kunz}}}, \bibnamefont{and}
  \bibinfo{author}{\bibfnamefont{S.}~\bibnamefont{{Mojica}}},
  \bibinfo{journal}{\prd} \textbf{\bibinfo{volume}{90}}, \bibinfo{eid}{061501}
  (\bibinfo{year}{2014}), \eprint{1407.6884}.

\bibitem[{\citenamefont{{Pani} et~al.}(2011{\natexlab{b}})\citenamefont{{Pani},
  {Berti}, {Cardoso}, and {Read}}}]{2011PhRvD..84j4035P}
\bibinfo{author}{\bibfnamefont{P.}~\bibnamefont{{Pani}}},
  \bibinfo{author}{\bibfnamefont{E.}~\bibnamefont{{Berti}}},
  \bibinfo{author}{\bibfnamefont{V.}~\bibnamefont{{Cardoso}}},
  \bibnamefont{and} \bibinfo{author}{\bibfnamefont{J.}~\bibnamefont{{Read}}},
  \bibinfo{journal}{\prd} \textbf{\bibinfo{volume}{84}}, \bibinfo{eid}{104035}
  (\bibinfo{year}{2011}{\natexlab{b}}), \eprint{1109.0928}.

\bibitem[{\citenamefont{{Yagi} and {Yunes}}(2013)}]{2013Sci...341..365Y}
\bibinfo{author}{\bibfnamefont{K.}~\bibnamefont{{Yagi}}} \bibnamefont{and}
  \bibinfo{author}{\bibfnamefont{N.}~\bibnamefont{{Yunes}}},
  \bibinfo{journal}{Science} \textbf{\bibinfo{volume}{341}},
  \bibinfo{pages}{365} (\bibinfo{year}{2013}), \eprint{1302.4499}.

\bibitem[{\citenamefont{{Kokkotas} et~al.}(2001)\citenamefont{{Kokkotas},
  {Apostolatos}, and {Andersson}}}]{2001MNRAS.320..307K}
\bibinfo{author}{\bibfnamefont{K.~D.} \bibnamefont{{Kokkotas}}},
  \bibinfo{author}{\bibfnamefont{T.~A.} \bibnamefont{{Apostolatos}}},
  \bibnamefont{and}
  \bibinfo{author}{\bibfnamefont{N.}~\bibnamefont{{Andersson}}},
  \bibinfo{journal}{\mnras} \textbf{\bibinfo{volume}{320}},
  \bibinfo{pages}{307} (\bibinfo{year}{2001}), \eprint{gr-qc/9901072}.

\bibitem[{\citenamefont{{Benhar} et~al.}(2004)\citenamefont{{Benhar},
  {Ferrari}, and {Gualtieri}}}]{2004PhRvD..70l4015B}
\bibinfo{author}{\bibfnamefont{O.}~\bibnamefont{{Benhar}}},
  \bibinfo{author}{\bibfnamefont{V.}~\bibnamefont{{Ferrari}}},
  \bibnamefont{and}
  \bibinfo{author}{\bibfnamefont{L.}~\bibnamefont{{Gualtieri}}},
  \bibinfo{journal}{\prd} \textbf{\bibinfo{volume}{70}}, \bibinfo{eid}{124015}
  (\bibinfo{year}{2004}), \eprint{astro-ph/0407529}.

\bibitem[{\citenamefont{{Bl{\'a}zquez-Salcedo}
  et~al.}(2013)\citenamefont{{Bl{\'a}zquez-Salcedo}, {Gonz{\'a}lez-Romero}, and
  {Navarro-L{\'e}rida}}}]{2013PhRvD..87j4042B}
\bibinfo{author}{\bibfnamefont{J.~L.} \bibnamefont{{Bl{\'a}zquez-Salcedo}}},
  \bibinfo{author}{\bibfnamefont{L.~M.} \bibnamefont{{Gonz{\'a}lez-Romero}}},
  \bibnamefont{and}
  \bibinfo{author}{\bibfnamefont{F.}~\bibnamefont{{Navarro-L{\'e}rida}}},
  \bibinfo{journal}{\prd} \textbf{\bibinfo{volume}{87}}, \bibinfo{eid}{104042}
  (\bibinfo{year}{2013}), \eprint{1207.4651}.

\bibitem[{\citenamefont{{Bl{\'a}zquez-Salcedo}
  et~al.}(2014)\citenamefont{{Bl{\'a}zquez-Salcedo}, {Gonz{\'a}lez-Romero}, and
  {Navarro-L{\'e}rida}}}]{2014PhRvD..89d4006B}
\bibinfo{author}{\bibfnamefont{J.~L.} \bibnamefont{{Bl{\'a}zquez-Salcedo}}},
  \bibinfo{author}{\bibfnamefont{L.~M.} \bibnamefont{{Gonz{\'a}lez-Romero}}},
  \bibnamefont{and}
  \bibinfo{author}{\bibfnamefont{F.}~\bibnamefont{{Navarro-L{\'e}rida}}},
  \bibinfo{journal}{\prd} \textbf{\bibinfo{volume}{89}}, \bibinfo{eid}{044006}
  (\bibinfo{year}{2014}), \eprint{1307.1063}.

\bibitem[{\citenamefont{{Yagi}}(2012)}]{2012PhRvD..86h1504Y}
\bibinfo{author}{\bibfnamefont{K.}~\bibnamefont{{Yagi}}},
  \bibinfo{journal}{\prd} \textbf{\bibinfo{volume}{86}}, \bibinfo{eid}{081504}
  (\bibinfo{year}{2012}), \eprint{1204.4524}.

\bibitem[{\citenamefont{Regge and Wheeler}(1957)}]{PhysRev.108.1063}
\bibinfo{author}{\bibfnamefont{T.}~\bibnamefont{Regge}} \bibnamefont{and}
  \bibinfo{author}{\bibfnamefont{J.~A.} \bibnamefont{Wheeler}},
  \bibinfo{journal}{Phys. Rev.} \textbf{\bibinfo{volume}{108}},
  \bibinfo{pages}{1063} (\bibinfo{year}{1957}),
  \urlprefix\url{http://link.aps.org/doi/10.1103/PhysRev.108.1063}.

\bibitem[{\citenamefont{Zerilli}(1970)}]{PhysRevLett.24.737}
\bibinfo{author}{\bibfnamefont{F.~J.} \bibnamefont{Zerilli}},
  \bibinfo{journal}{Phys. Rev. Lett.} \textbf{\bibinfo{volume}{24}},
  \bibinfo{pages}{737} (\bibinfo{year}{1970}),
  \urlprefix\url{http://link.aps.org/doi/10.1103/PhysRevLett.24.737}.

\bibitem[{\citenamefont{Thorne}(1980)}]{RevModPhys.52.299}
\bibinfo{author}{\bibfnamefont{K.~S.} \bibnamefont{Thorne}},
  \bibinfo{journal}{Rev. Mod. Phys.} \textbf{\bibinfo{volume}{52}},
  \bibinfo{pages}{299} (\bibinfo{year}{1980}),
  \urlprefix\url{http://link.aps.org/doi/10.1103/RevModPhys.52.299}.

\bibitem[{\citenamefont{Ascher et~al.}(1979)\citenamefont{Ascher, Christiansen,
  and Russell}}]{colsys1979}
\bibinfo{author}{\bibfnamefont{U.}~\bibnamefont{Ascher}},
  \bibinfo{author}{\bibfnamefont{J.}~\bibnamefont{Christiansen}},
  \bibnamefont{and} \bibinfo{author}{\bibfnamefont{R.~D.}
  \bibnamefont{Russell}}, \bibinfo{journal}{Mathematics of Computation}
  \textbf{\bibinfo{volume}{33}}, \bibinfo{pages}{pp. 659}
  (\bibinfo{year}{1979}).

\bibitem[{\citenamefont{Israel}(1966)}]{israel66}
\bibinfo{author}{\bibfnamefont{W.}~\bibnamefont{Israel}}, \bibinfo{journal}{Il
  Nuovo Cimento B Series 10} \textbf{\bibinfo{volume}{44}}, \bibinfo{pages}{1}
  (\bibinfo{year}{1966}), ISSN \bibinfo{issn}{0369-3554},
  \urlprefix\url{http://dx.doi.org/10.1007/BF02710419}.

\bibitem[{\citenamefont{{Douchin} and {Haensel}}(2001)}]{Douchin2001}
\bibinfo{author}{\bibfnamefont{F.}~\bibnamefont{{Douchin}}} \bibnamefont{and}
  \bibinfo{author}{\bibfnamefont{P.}~\bibnamefont{{Haensel}}},
  \bibinfo{journal}{\aap} \textbf{\bibinfo{volume}{380}}, \bibinfo{pages}{151}
  (\bibinfo{year}{2001}), \eprint{astro-ph/0111092}.

\bibitem[{\citenamefont{Akmal et~al.}(1998)\citenamefont{Akmal, Pandharipande,
  and Ravenhall}}]{PhysRevC.58.1804}
\bibinfo{author}{\bibfnamefont{A.}~\bibnamefont{Akmal}},
  \bibinfo{author}{\bibfnamefont{V.~R.} \bibnamefont{Pandharipande}},
  \bibnamefont{and} \bibinfo{author}{\bibfnamefont{D.~G.}
  \bibnamefont{Ravenhall}}, \bibinfo{journal}{Phys. Rev. C}
  \textbf{\bibinfo{volume}{58}}, \bibinfo{pages}{1804} (\bibinfo{year}{1998}),
  \urlprefix\url{http://link.aps.org/doi/10.1103/PhysRevC.58.1804}.

\bibitem[{\citenamefont{{Bednarek} et~al.}(2012)\citenamefont{{Bednarek},
  {Haensel}, {Zdunik}, {Bejger}, and {Ma{\'n}ka}}}]{Bednarek2012}
\bibinfo{author}{\bibfnamefont{I.}~\bibnamefont{{Bednarek}}},
  \bibinfo{author}{\bibfnamefont{P.}~\bibnamefont{{Haensel}}},
  \bibinfo{author}{\bibfnamefont{J.~L.} \bibnamefont{{Zdunik}}},
  \bibinfo{author}{\bibfnamefont{M.}~\bibnamefont{{Bejger}}}, \bibnamefont{and}
  \bibinfo{author}{\bibfnamefont{R.}~\bibnamefont{{Ma{\'n}ka}}},
  \bibinfo{journal}{\aap} \textbf{\bibinfo{volume}{543}}, \bibinfo{eid}{A157}
  (\bibinfo{year}{2012}), \eprint{1111.6942}.

\bibitem[{\citenamefont{Weissenborn et~al.}(2012)\citenamefont{Weissenborn,
  Chatterjee, and Schaffner-Bielich}}]{PhysRevC.85.065802}
\bibinfo{author}{\bibfnamefont{S.}~\bibnamefont{Weissenborn}},
  \bibinfo{author}{\bibfnamefont{D.}~\bibnamefont{Chatterjee}},
  \bibnamefont{and}
  \bibinfo{author}{\bibfnamefont{J.}~\bibnamefont{Schaffner-Bielich}},
  \bibinfo{journal}{Phys. Rev. C} \textbf{\bibinfo{volume}{85}},
  \bibinfo{pages}{065802} (\bibinfo{year}{2012}),
  \urlprefix\url{http://link.aps.org/doi/10.1103/PhysRevC.85.065802}.

\bibitem[{\citenamefont{Weissenborn et~al.}(2014)\citenamefont{Weissenborn,
  Chatterjee, and Schaffner-Bielich}}]{PhysRevC.90.019904}
\bibinfo{author}{\bibfnamefont{S.}~\bibnamefont{Weissenborn}},
  \bibinfo{author}{\bibfnamefont{D.}~\bibnamefont{Chatterjee}},
  \bibnamefont{and}
  \bibinfo{author}{\bibfnamefont{J.}~\bibnamefont{Schaffner-Bielich}},
  \bibinfo{journal}{Phys. Rev. C} \textbf{\bibinfo{volume}{90}},
  \bibinfo{pages}{019904} (\bibinfo{year}{2014}),
  \urlprefix\url{http://link.aps.org/doi/10.1103/PhysRevC.90.019904}.

\bibitem[{\citenamefont{{Alford} et~al.}(2005)\citenamefont{{Alford}, {Braby},
  {Paris}, and {Reddy}}}]{2005ApJ...629..969A}
\bibinfo{author}{\bibfnamefont{M.}~\bibnamefont{{Alford}}},
  \bibinfo{author}{\bibfnamefont{M.}~\bibnamefont{{Braby}}},
  \bibinfo{author}{\bibfnamefont{M.}~\bibnamefont{{Paris}}}, \bibnamefont{and}
  \bibinfo{author}{\bibfnamefont{S.}~\bibnamefont{{Reddy}}},
  \bibinfo{journal}{\apj} \textbf{\bibinfo{volume}{629}}, \bibinfo{pages}{969}
  (\bibinfo{year}{2005}), \eprint{nucl-th/0411016}.

\bibitem[{\citenamefont{{Bonanno} and {Sedrakian}}(2012)}]{Sed2012}
\bibinfo{author}{\bibfnamefont{L.}~\bibnamefont{{Bonanno}}} \bibnamefont{and}
  \bibinfo{author}{\bibfnamefont{A.}~\bibnamefont{{Sedrakian}}},
  \bibinfo{journal}{\aap} \textbf{\bibinfo{volume}{539}}, \bibinfo{eid}{A16}
  (\bibinfo{year}{2012}), \eprint{1108.0559}.

\bibitem[{\citenamefont{{Weissenborn} et~al.}(2011)\citenamefont{{Weissenborn},
  {Sagert}, {Pagliara}, {Hempel}, and
  {Schaffner-Bielich}}}]{2011ApJ...740L..14W}
\bibinfo{author}{\bibfnamefont{S.}~\bibnamefont{{Weissenborn}}},
  \bibinfo{author}{\bibfnamefont{I.}~\bibnamefont{{Sagert}}},
  \bibinfo{author}{\bibfnamefont{G.}~\bibnamefont{{Pagliara}}},
  \bibinfo{author}{\bibfnamefont{M.}~\bibnamefont{{Hempel}}}, \bibnamefont{and}
  \bibinfo{author}{\bibfnamefont{J.}~\bibnamefont{{Schaffner-Bielich}}},
  \bibinfo{journal}{\apjl} \textbf{\bibinfo{volume}{740}}, \bibinfo{eid}{L14}
  (\bibinfo{year}{2011}), \eprint{1102.2869}.

\bibitem[{\citenamefont{Lenzi et~al.}(2009)\citenamefont{Lenzi, Malheiro,
  Marinho, Marranghello, and Providência}}]{1742-6596-154-1-012039}
\bibinfo{author}{\bibfnamefont{C.~H.} \bibnamefont{Lenzi}},
  \bibinfo{author}{\bibfnamefont{M.}~\bibnamefont{Malheiro}},
  \bibinfo{author}{\bibfnamefont{R.~M.} \bibnamefont{Marinho}},
  \bibinfo{author}{\bibfnamefont{G.~F.} \bibnamefont{Marranghello}},
  \bibnamefont{and}
  \bibinfo{author}{\bibfnamefont{C.}~\bibnamefont{Providência}},
  \bibinfo{journal}{Journal of Physics: Conference Series}
  \textbf{\bibinfo{volume}{154}}, \bibinfo{pages}{012039}
  (\bibinfo{year}{2009}),
  \urlprefix\url{http://stacks.iop.org/1742-6596/154/i=1/a=012039}.

\bibitem[{\citenamefont{Wen et~al.}(2009)\citenamefont{Wen, Li, and
  Krastev}}]{PhysRevC.80.025801}
\bibinfo{author}{\bibfnamefont{D.-H.} \bibnamefont{Wen}},
  \bibinfo{author}{\bibfnamefont{B.-A.} \bibnamefont{Li}}, \bibnamefont{and}
  \bibinfo{author}{\bibfnamefont{P.~G.} \bibnamefont{Krastev}},
  \bibinfo{journal}{Phys. Rev. C} \textbf{\bibinfo{volume}{80}},
  \bibinfo{pages}{025801} (\bibinfo{year}{2009}),
  \urlprefix\url{http://link.aps.org/doi/10.1103/PhysRevC.80.025801}.

\bibitem[{\citenamefont{Kokkotas and Schutz}(1986)}]{4727}
\bibinfo{author}{\bibfnamefont{K.~D.} \bibnamefont{Kokkotas}} \bibnamefont{and}
  \bibinfo{author}{\bibfnamefont{B.~F.} \bibnamefont{Schutz}},
  \bibinfo{journal}{Gen. Relativ. Gravit.} \textbf{\bibinfo{volume}{18}},
  \bibinfo{pages}{913} (\bibinfo{year}{1986}).

\bibitem[{\citenamefont{Kokkotas and Schutz}(1992)}]{532}
\bibinfo{author}{\bibfnamefont{K.~D.} \bibnamefont{Kokkotas}} \bibnamefont{and}
  \bibinfo{author}{\bibfnamefont{B.~F.} \bibnamefont{Schutz}},
  \bibinfo{journal}{Mon. Not. R. Astron. Soc.} \textbf{\bibinfo{volume}{255}},
  \bibinfo{pages}{119} (\bibinfo{year}{1992}).

\bibitem[{\citenamefont{Benhar et~al.}(2004)\citenamefont{Benhar, Ferrari, and
  Gualtieri}}]{PhysRevD.70.124015}
\bibinfo{author}{\bibfnamefont{O.}~\bibnamefont{Benhar}},
  \bibinfo{author}{\bibfnamefont{V.}~\bibnamefont{Ferrari}}, \bibnamefont{and}
  \bibinfo{author}{\bibfnamefont{L.}~\bibnamefont{Gualtieri}},
  \bibinfo{journal}{Phys. Rev. D} \textbf{\bibinfo{volume}{70}},
  \bibinfo{pages}{124015} (\bibinfo{year}{2004}),
  \urlprefix\url{http://link.aps.org/doi/10.1103/PhysRevD.70.124015}.

\bibitem[{\citenamefont{Andersson and Kokkotas}(1996)}]{PhysRevLett.77.4134}
\bibinfo{author}{\bibfnamefont{N.}~\bibnamefont{Andersson}} \bibnamefont{and}
  \bibinfo{author}{\bibfnamefont{K.~D.} \bibnamefont{Kokkotas}},
  \bibinfo{journal}{Phys. Rev. Lett.} \textbf{\bibinfo{volume}{77}},
  \bibinfo{pages}{4134} (\bibinfo{year}{1996}),
  \urlprefix\url{http://link.aps.org/doi/10.1103/PhysRevLett.77.4134}.

\bibitem[{\citenamefont{Andersson and Kokkotas}(1998)}]{Andersson01101998}
\bibinfo{author}{\bibfnamefont{N.}~\bibnamefont{Andersson}} \bibnamefont{and}
  \bibinfo{author}{\bibfnamefont{K.~D.} \bibnamefont{Kokkotas}},
  \bibinfo{journal}{Monthly Notices of the Royal Astronomical Society}
  \textbf{\bibinfo{volume}{299}}, \bibinfo{pages}{1059} (\bibinfo{year}{1998}),
  \eprint{http://mnras.oxfordjournals.org/content/299/4/1059.full.pdf+html},
  \urlprefix\url{http://mnras.oxfordjournals.org/content/299/4/1059.abstract}.

\end{thebibliography}

\end{document}